\documentclass[amstex,fleqn,thmsa,a4,12pt]{article}
\usepackage{amsmath}

\input{tcilatex}
\input tcilatex

\begin{document}

\begin{titlepage}
\setcounter{page}{1}

\title{POINCARE ALGEBRA AND SPACE-TIME CRITICAL DIMENSIONS FOR PARABOSONIC STRINGS%
\thanks{%
This work was supported by the Algerian Ministry of Education and Research
under contract No.D2501/14/2000.}}
\author{N.BELALOUI\thanks{E-mail : n.belaloui@wissal.dz}, AND H.BENNACER \\
LPMPS, D\'{e}partement de Physique, Facult\'{e} des Sciences,\\
Universit\'{e} Mentouri Constantine, \\
Constantine, Algeria.}
\maketitle

\begin{abstract}
We construct the parabosonic string formalism based on the paraquantization
of both the center of mass variables and the excitation modes of the
string. A critical study of the different commutators of the Poincar\'{e}
algebra based on the redefinition of its generators and the direct treatment
using trilinear relations is done.\ Space-time critical dimensions $D$ as
functions of the paraquantization order $Q$ are obtained
\end{abstract}
\end{titlepage}

\section{Introduction}

\setcounter{page}{2}

As far back as 1950,Wigner [1] demonstrated that, for satisfying the wave
particle duality, which is a direct consequence of the Heisenberg equations
of motion, the set of canonical commutation relations $\left[ q_{i},p_{j}%
\right] =\imath \hbar \delta _{ij},\ $ $\left[ q_{i},q_{j}\right] =\left[
p_{i},p_{j}\right] =0$ (which correspond to the usual procedure of canonical
quantization ) is a possible solution, but is by no means the only one.
Paraquantization, as a generalization of the quantization, was first
introduced by Green [2] . Indeed, one basing itself on trilinear commutation
relations, paraqantization consists of a generalization of
creation-annihilation operators algebra for bosons and fermions. We note
also that the paraquantization is characterized by a parameter $Q$, the
order of paraquantization, such that $Q=1$ corresponds to the ordinary
quantization. The details of these questions can be found in [3].This work
consists in doing a critical study of Poincar\'{e} algebra in the case of
parabosonic strings. To set the notations, we begin with a brief summary of
some familiar results in bosonic string theory.

The action is postulated as [4],[5] .\bigskip 
\begin{equation}
S=-\int_{\tau _{0}}^{\tau _{1}}d\tau \int_{0}^{\pi }d\sigma L  \tag{1}
\end{equation}
with the lagrangian 
\begin{equation}
L=\frac{1}{2\pi \alpha ^{\prime }}\sqrt{\left( \overset{.}{X}X^{\prime
}\right) ^{2}-\overset{.}{X^{2}}X^{\prime 2}}  \tag{2}
\end{equation}
where 
\begin{eqnarray}
\overset{.}{X^{\mu }} &=&\frac{\partial X^{\mu }\left( \sigma ,\tau \right) 
}{\partial \tau }  \TCItag{3} \\
X^{\prime \mu } &=&\frac{\partial X^{\mu }\left( \sigma ,\tau \right) }{%
\partial \sigma }  \TCItag{4}
\end{eqnarray}
$\tau $ is a time like evolution parameter, while the parameter $\sigma $
labels points on the string . Notice that the conjugate momentum to $X^{\mu
}\left( \sigma ,\tau \right) $ is $\mathcal{P}^{\mu }\left( \sigma ,\tau
\right) =-\displaystyle\frac{\partial L}{\partial \overset{.}{X}_{\mu }}%
\qquad $

In the orthonormal gauge, the equations of motion become linear. The
solutions are: 
\begin{equation}
X^{\mu }\left( \sigma ,\tau \right) =x^{\mu }+p^{\mu }\tau +\sum_{n\neq 0}%
\displaystyle\frac{1}{n}\alpha _{n}^{\mu }\left( 0\right) \exp \left(
-\imath n\tau \right) \cos n\sigma  \tag{5}
\end{equation}
where $x^{\mu }$ and $p^{\mu }$ are respectively the ''center of mass''
coordinates and the total energy momentum of the string. The total angular
momentum of the string is given by 
\begin{equation}
M^{\mu \nu }=x^{\mu }p^{\nu }-x^{\nu }p^{\mu }-\imath \sum_{n=1}^{\infty }%
\frac{1}{n}\left( \alpha _{-n}^{\mu }\alpha _{n}^{\nu }-\alpha _{-n}^{\nu
}\alpha _{n}^{\mu }\right)  \tag{6}
\end{equation}

A first study of the paraquantum Poincar\'{e} algebra was done by F.Ardalan
and F.Mansouri [6] . This study is based on the particular manner in which
the center of mass variables of the string are to be handled. Indeed, these
authors impose on the center of mass coordinates and the total energy
momentum operators of the string to satisfy ordinary commutation relations.\
This is done by the choice of a specific direction in the paraspace of the
Green components (see (19)). This requires relative paracommutation
relations between the center of mass coordinates and the excitation modes of
the string.\ In this hypothesis, they find that the resulting theory is
Poincar\'{e} invariant if the dimension $D$ of the space-time and the order $%
Q$ of the paraquantization are related by the expression $D=2+\frac{24}{Q}.$

Notice that , with the same hypothesis, this relation has been found by
other methods [7], [8].

This work consist in paraquantizing the theory by reinterpreting the
classical string variables $X^{\mu }\left( \sigma ,\tau \right) \mathcal{\ }$%
and $\mathcal{P}^{\mu }\left( \sigma ,\tau \right) $ as operators satisfying
the paraquantum relations (7-10) ( subsection 2 .1). This is done by
requiring that both the center of mass variables and the excitation modes of
the string verify paraquantum relations (11-18) ( subsection 2 .1). The
first commutator of the Poincar\'{e} algebra ( $\left[ p^{\mu },p^{\nu }%
\right] =0\,!$) means that the operators $p^{\mu }$ obey to bilinear
commutation relations which is against the paracommutation relations based
on trilinear relations as for example $\left[ p^{\mu },\left[ p^{\nu
},p^{\sigma }\right] _{+}\right] =0$ . Nevertheless , whith the only use of
triliner relations we prove that the two others commutators of the algebra
are satisfyied and that the relation between the space-time dimension $D$
and the order of the paraquantization $Q$ is still $D=2+\frac{24}{Q}.$

\section{Paraquantum formalism of bosonic strings}

\subsection{Covariant gauge}

The paraquantization of the theory is carried out by reinterpreting the
classical dynamical variables $X^{\mu }\left( \sigma ,\tau \right) $ and $%
P^{\mu }\left( \sigma ,\tau \right) $ as operators satisfying the so-called
trilinear commutation relations\ 
\begin{equation}
\!\!\hspace*{-4mm}\left[ X^{\mu }\left( \sigma ,\tau \right) ,\left[ P^{\nu
}\left( \sigma ^{\prime },\tau \right) ,P^{\rho }\left( \sigma ^{\prime
\prime },\tau \right) \right] _{+}\right] =2\imath g^{\mu \nu }P^{\rho
}\delta \left( \sigma -\sigma ^{\prime }\right) +2\imath g^{\mu \rho }P^{\nu
}\delta \left( \sigma -\sigma ^{\prime \prime }\right) \quad   \tag{7}
\end{equation}
\begin{equation}
\hspace*{-5mm}\left[ P^{\mu }\left( \sigma ,\tau \right) ,\left[ X^{\nu
}\left( \sigma ^{\prime },\tau \right) ,X^{\rho }\left( \sigma ^{\prime
\prime },\tau \right) \right] _{+}\right] =-2\imath g^{\mu \nu }X^{\rho
}\delta \left( \sigma -\sigma ^{\prime }\right) -2\imath g^{\mu \rho }X^{\nu
}\delta \left( \sigma -\sigma ^{\prime \prime }\right)   \tag{8}
\end{equation}
\begin{equation}
\left[ X^{\mu }\left( \sigma ,\tau \right) ,\left[ X^{\nu }\left( \sigma
^{\prime },\tau \right) ,P^{\rho }\left( \sigma ^{\prime \prime },\tau
\right) \right] _{+}\right] =2\imath g^{\mu \rho }X^{\nu }\delta \left(
\sigma -\sigma ^{\prime \prime }\right)   \tag{9}
\end{equation}
\begin{equation}
\left[ P^{\mu }\left( \sigma ,\tau \right) ,\left[ X^{\nu }\left( \sigma
^{\prime },\tau \right) ,P^{\rho }\left( \sigma ^{\prime \prime },\tau
\right) \right] _{+}\right] =2\imath g^{\mu \nu }P^{\rho }\delta \left(
\sigma -\sigma ^{\prime }\right)   \tag{10}
\end{equation}
Rewritten in terms of $x^{\mu }$, $p^{\mu }$ and $\alpha _{n}^{\mu }$
defined by (5) , equations (7-10) are equivalent to : 
\begin{eqnarray}
\left[ x^{\mu },\left[ x^{\nu },p^{\rho }\right] _{+}\right]  &=&2\imath
g^{\mu \rho }x^{\nu }  \TCItag{11} \\
\left[ x^{\mu },\left[ p^{\nu },p^{\rho }\right] _{+}\right]  &=&2\imath
\left( g^{\mu \nu }p^{\rho }+g^{\mu \rho }p^{\nu }\right)   \TCItag{12} \\
\left[ x^{\mu },\left[ p^{\nu },\alpha _{n}^{\rho }\right] _{+}\right] 
&=&2\imath g^{\mu \nu }\alpha _{n}^{\rho }  \TCItag{13} \\
\left[ p^{\mu },\left[ x^{\nu },p^{\rho }\right] _{+}\right]  &=&-2\imath
g^{\mu \nu }p^{\rho }  \TCItag{14} \\
\left[ p^{\mu },\left[ x^{\nu },x^{\rho }\right] _{+}\right]  &=&-2\imath
\left( g^{\mu \nu }x^{\rho }+g^{\mu \rho }x^{\nu }\right)   \TCItag{15} \\
\left[ \alpha _{n}^{\mu },\left[ \alpha _{m}^{\nu },\alpha _{l}^{\rho }%
\right] _{+}\right]  &=&2\left( g^{\mu \nu }n\delta _{n+m,0}\alpha
_{l}^{\rho }+g^{\mu \rho }n\delta _{n+l,0}\alpha _{m}^{\nu }\right)  
\TCItag{16} \\
\left[ \alpha _{n}^{\mu },\left[ p^{\nu },\alpha _{m}^{\rho }\right] _{+}%
\right]  &=&2ng^{\mu \rho }\delta _{n+m,0}p^{\nu }  \TCItag{17} \\
\left[ \alpha _{n}^{\mu },\left[ x^{\nu },\alpha _{m}^{\rho }\right] _{+}%
\right]  &=&2ng^{\mu \rho }\delta _{n+m,0}x^{\nu }  \TCItag{18}
\end{eqnarray}
and all the other commutators are null.

Next we introduce the Green decomposition of the operators $x^{\mu }$, $%
p^{\mu }$ and $\alpha _{n}^{\mu }$ defined by : 
\begin{equation}
x^{\mu }=\sum_{\alpha =1}^{Q}x^{\mu \left( \alpha \right) }\quad ;\quad
p^{\mu }=\sum_{\alpha =1}^{Q}p^{\mu \left( \alpha \right) }\quad ;\quad
\alpha _{n}^{\mu }=\sum_{\beta =1}^{Q}\alpha _{n}^{\mu \left( \beta \right) }
\tag{19}
\end{equation}
where Q is the order of paraquantization, such that the trilinear
commutation relations (7-10) transform to bilinear commutation relations of
an anomalous case 
\begin{eqnarray}
\left[ X^{\mu (\alpha )}\left( \sigma ,\tau \right) ,P^{\nu (\alpha
)}(\sigma ^{\prime },\tau )\right]  &=&\imath g^{\mu \nu }\delta (\sigma
-\sigma ^{\prime })  \notag \\
\left[ X^{\mu (\alpha )}\left( \sigma ,\tau \right) ,P^{\nu (\beta )}(\sigma
^{\prime },\tau )\right] _{+} &=&0\quad \quad \text{for }\alpha \neq \beta  
\notag \\
\left[ X^{\mu (\alpha )}\left( \sigma ,\tau \right) ,X^{\nu (\alpha
)}(\sigma ^{\prime },\tau )\right]  &=&\left[ P^{\mu (\alpha )}\left( \sigma
,\tau \right) ,P^{\nu (\alpha )}(\sigma ^{\prime },\tau )\right] =0 
\TCItag{20} \\
\left[ X^{\mu (\alpha )}\left( \sigma ,\tau \right) ,X^{\nu (\beta )}(\sigma
^{\prime },\tau )\right] _{+} &=&\left[ P^{\mu (\alpha )}\left( \sigma ,\tau
\right) ,P^{\nu (\beta _{{}})}(\sigma ^{\prime },\tau )\right] _{+}=0\quad
\quad \text{for }\alpha \neq \beta   \notag
\end{eqnarray}

In the same way for the relations (11-18) 
\begin{eqnarray}
\left[ x^{\mu \left( \sigma \right) },p^{\nu \left( \sigma \right) }\right] 
&=&\imath g^{\mu \nu }  \notag \\
\left[ x^{\mu \left( \sigma _{1}\right) },p^{\nu \left( \sigma _{2}\right) }%
\right] _{+} &=&0\quad \quad \quad \quad \sigma _{1}\neq \sigma _{2}  \notag
\\
\left[ p^{\mu \left( \sigma \right) },p^{\nu \left( \sigma \right) }\right] 
&=&\left[ x^{\mu \left( \sigma \right) },x^{\nu \left( \sigma \right) }%
\right] =0  \notag \\
\left[ p^{\mu \left( \sigma _{1}\right) },p^{\nu \left( \sigma _{2}\right) }%
\right] _{+} &=&\left[ x^{\mu \left( \sigma _{1}\right) },x^{\nu \left(
\sigma _{2}\right) }\right] _{+}=0\quad \quad \quad \quad \sigma _{1}\neq
\sigma _{2}  \notag \\
\left[ \alpha _{n}^{\mu \left( \sigma \right) },\alpha _{m}^{\nu \left(
\sigma \right) }\right]  &=&ng^{\mu \nu }\delta _{n+m,0}  \TCItag{21} \\
\left[ \alpha _{n}^{\mu \left( \sigma _{1}\right) },\alpha _{l}^{\nu \left(
\sigma _{2}\right) }\right] _{+} &=&0\quad \quad \quad \quad \sigma _{1}\neq
\sigma _{2}  \notag \\
\left[ x^{\mu \left( \sigma \right) },\alpha _{n}^{\nu \left( \sigma \right)
}\right]  &=&\left[ p^{\mu \left( \sigma \right) },\alpha _{n}^{\nu \left(
\sigma \right) }\right] =0  \notag \\
\left[ x^{\mu \left( \sigma _{1}\right) },\alpha _{n}^{\nu \left( \sigma
_{2}\right) }\right] _{+} &=&\left[ p^{\mu \left( \sigma _{1}\right)
},\alpha _{n}^{\nu \left( \sigma _{2}\right) }\right] _{+}=0\quad \quad
\quad \quad \sigma _{1}\neq \sigma _{2}  \notag
\end{eqnarray}

Notice here that in the Ardalan and Mansouri hypothesis [6], $\left[ X^{\mu
(\alpha )}\left( \sigma ,\tau \right) ,P^{\nu (\alpha )}(\sigma ^{\prime
},\tau )\right] =$ $\imath g^{\mu \nu }\left[ \delta (\sigma -\sigma
^{\prime })-(1-\delta _{\alpha 1})\right] $ which is not compatible with the
relations (7-10).

\subsection{Transverse gauge}

In the same way as before, paraquantizing the theory in this gauge comes
down to reinterprete the independant classical dynamical variables $%
x^{-},\,p^{+},\,x^{i},\,p^{i}$ and $\alpha _{n}^{i}$ as operators satisfying
the paracommutation relations : 
\begin{eqnarray}
\left[ x^{i},\left[ p^{j},p^{k}\right] _{+}\right]  &=&2\imath \left(
g^{ij}p^{k}+g^{ik}p^{j}\right)   \notag \\
\left[ \alpha _{n}^{i},\left[ \alpha _{m}^{j},\alpha _{l}^{k}\right] _{+}%
\right]  &=&2\left( g^{ij}n\delta _{n+m,0}\alpha _{l}^{k}+g^{ik}n\delta
_{n+l,0}\alpha _{m}^{j}\right)   \notag \\
\left[ x^{i},\left[ p^{j},A\right] _{+}\right]  &=&2\imath \delta ^{ij}A 
\TCItag{22} \\
\left[ x^{-},\left[ p^{+},B\right] _{+}\right]  &=&2\imath B  \notag \\
\left[ \alpha _{n}^{i},\left[ \alpha _{m}^{j},C\right] _{+}\right] 
&=&2n\delta _{n+m,0}\delta ^{ij}C  \notag
\end{eqnarray}
where $A$ ,$B$, and $C$ are given by :

$A=x^{-},\,p^{+},\,x^{k},\,$or $\alpha _{n}^{k}.$

$B=x^{-},\,x^{k},\,p^{k}$ or $\alpha _{n}^{k}.$

$C=x^{-},\,p^{+},\,x^{k}\,$or $p^{k}.$

Similarly, applying the Green decomposition

\begin{eqnarray}
x^{i} &=&\sum_{\alpha =1}^{Q}x^{i\left( \alpha \right) }\quad ;\quad
p^{i}=\sum_{\alpha =1}^{Q}p^{i\left( \alpha \right) }\quad ;\quad \alpha
_{n}^{i}=\sum_{\beta =1}^{Q}\alpha _{n}^{i\left( \beta \right) }  \notag \\
x^{-} &=&\sum_{\alpha =1}^{Q}x^{-\left( \alpha \right) }\quad ;\quad
p^{+}=\sum_{\alpha =1}^{Q}p^{+\left( \alpha \right) }  \TCItag{23}
\end{eqnarray}
the set of equations (22) is equivalent to the bilinear relations

\begin{eqnarray}
\left[ x^{i\left( \alpha \right) },p^{j\left( \alpha \right) }\right] 
&=&\imath \delta ^{ij}\quad \quad ;\quad \quad \left[ x^{i\left( \alpha
\right) },p^{j\left( \beta \right) }\right] _{+}=0\quad \alpha \neq \beta  
\notag \\
\left[ x^{-\left( \alpha \right) },p^{+\left( \alpha \right) }\right] 
&=&\imath \quad \quad ;\quad \quad \left[ x^{-\left( \alpha \right)
},p^{+\left( \beta \right) }\right] _{+}=0\quad \alpha \neq \beta  
\TCItag{24} \\
\left[ \alpha _{n}^{i\left( \alpha \right) },\alpha _{m}^{j\left( \alpha
\right) }\right]  &=&n\delta ^{ij}\delta _{n+m,0}\quad \quad ;\quad \quad %
\left[ \alpha _{n}^{i\left( \alpha \right) },\alpha _{m}^{j\left( \beta
\right) }\right] _{+}=0\quad \alpha \neq \beta   \notag
\end{eqnarray}
and all the other commutators (and anticommutators) of the type $\left[
A^{\left( \alpha \right) },B^{\left( \alpha \right) }\right] =0$ (and $\left[
A^{\left( \alpha \right) },B^{\left( \beta \right) }\right] _{+}=0$, for $%
\alpha \neq \beta $).

\section{ Paraquantum Poincar\'{e} algebra}

\subsection{Introduction}

Just as an example, we beging by showing that, if we take the classical form
of the Poincar\'{e} algebra generators in terms of operators (like in the
ordinary case ), the direct use of the trilinear paracommutation relations
is not possible and, of course, the algebra is completely violated .These
generators are then written in the form

\begin{equation*}
M^{\mu \nu }=l^{\mu \nu }+E^{\mu \nu }.
\end{equation*}
where 
\begin{equation}
l^{\mu \nu }=\sum_{\alpha ,\beta =1}^{Q}\left[ x^{\mu \left( \alpha \right)
}p^{\nu \left( \beta \right) }-x^{\nu \left( \alpha \right) }p^{\mu \left(
\beta \right) }\right]   \tag{25}
\end{equation}
and 
\begin{equation}
E^{\mu \nu }=\sum_{\alpha ,\beta =1}^{Q}\left[ -\imath \sum_{n=1}^{+\infty }%
\frac{1}{n}\left( \alpha _{-n}^{\mu \left( \alpha \right) }\alpha _{n}^{\nu
\left( \beta \right) }-\alpha _{-n}^{\nu \left( \alpha \right) }\alpha
_{n}^{\mu \left( \beta \right) }\right) \right]   \tag{26}
\end{equation}

Let us consider the first commutator $\left[ p^{\mu },p^{\nu }\right] .$ It
is an easy matter to perform this commutator which gives

\begin{equation}
\left[ p^{\mu },p^{\nu }\right] =2\sum_{\alpha \neq \beta }p^{\mu \left(
\alpha \right) }p^{\nu \left( \beta \right) }\neq 0\,!  \tag{27}
\end{equation}

We now perform the second commutator of the algebra 
\begin{equation}
\left[ p^{\mu },M^{\alpha \beta }\right] =\left[ p^{\mu },l^{\alpha \beta }%
\right] +\left[ p^{\mu },E^{\alpha \beta }\right]   \tag{28}
\end{equation}

Taking the Green decomposition with the use of (21), the first term gives 
\begin{multline}
\left[ p^{\mu },l^{\alpha \beta }\right] =-\imath Q\left( g^{\mu \beta
}p^{\alpha }-g^{\mu \alpha }p^{\beta }\right) +2\sum_{\sigma _{1}\neq \sigma
_{2}}^{Q}\left( p^{\mu \left( \sigma _{1}\right) }x^{\alpha \left( \sigma
_{2}\right) }p^{\beta }-p^{\mu \left( \sigma _{1}\right) }x^{\beta \left(
\sigma _{2}\right) }p^{\mu }\right.   \notag \\
\left. +x^{\alpha }p^{\mu \left( \sigma _{1}\right) }p^{\beta \left( \sigma
_{2}\right) }-x^{\beta }p^{\mu \left( \sigma _{1}\right) }p^{\alpha \left(
\sigma _{2}\right) }\right)   \tag{29}
\end{multline}

Similarly for the second term, we obtain 
\begin{multline}
\left[ p^{\mu },E^{\alpha \beta }\right] =-2\imath \sum_{n=1}^{+\infty }%
\frac{1}{n}\sum_{\sigma _{1}\neq \sigma _{2}}\left( p^{\mu \left( \sigma
_{1}\right) }\alpha _{-n}^{\alpha \left( \sigma _{2}\right) }\alpha
_{n}^{\beta }-p^{\mu \left( \sigma _{1}\right) }\alpha _{-n}^{\beta \left(
\sigma _{2}\right) }\alpha _{n}^{\alpha }\right.   \notag \\
\left. +\alpha _{-n}^{\alpha }p^{\mu \left( \sigma _{1}\right) }\alpha
_{n}^{\beta \left( \sigma _{2}\right) }-\alpha _{-n}^{\beta }p^{\mu \left(
\sigma _{1}\right) }\alpha _{n}^{\alpha \left( \sigma _{2}\right) }\right)  
\tag{30}
\end{multline}

When combined with (29) and (30), (28) gives 
\begin{multline}
\left[ p^{\mu },M^{\alpha \beta }\right] =-\imath Q\left( g^{\mu \beta
}p^{\alpha }-g^{\mu \alpha }p^{\beta }\right)   \notag \\
-2\sum_{\sigma _{1}\neq \sigma _{2}}\left\{ p^{\mu \left( \sigma _{1}\right)
}x^{\alpha \left( \sigma _{2}\right) }p^{\beta }-p^{\mu \left( \sigma
_{1}\right) }x^{\beta \left( \sigma _{2}\right) }p^{\alpha }+x^{\alpha
}p^{\mu \left( \sigma _{1}\right) }p^{\beta \left( \sigma _{2}\right)
}-x^{\beta }p^{\mu \left( \sigma _{1}\right) }p^{\alpha \left( \sigma
_{2}\right) }\right.   \notag \\
\left. -\imath \sum_{n=1}^{+\infty }\frac{1}{n}\left( p^{\mu \left( \sigma
_{1}\right) }\alpha _{-n}^{\alpha \left( \sigma _{2}\right) }\alpha
_{n}^{\beta }-p^{\mu \left( \sigma _{1}\right) }\alpha _{-n}^{\beta \left(
\sigma _{2}\right) }\alpha _{n}^{\alpha }+\alpha _{-n}^{\alpha }p^{\mu
\left( \sigma _{1}\right) }\alpha _{n}^{\beta \left( \sigma _{2}\right)
}-\alpha _{-n}^{\beta }p^{\mu \left( \sigma _{1}\right) }\alpha _{n}^{\alpha
\left( \sigma _{2}\right) }\right) \right\}   \tag{31}
\end{multline}

The third commutator of the algebra can be performed as follows : 
\begin{equation}
\left[ M^{\mu \nu },M^{\alpha \beta }\right] =\left[ l^{\mu \nu },l^{\alpha
\beta }\right] +\left( \left[ l^{\mu \nu },E^{\alpha \beta }\right] -\left(
\mu \leftrightarrow \alpha ,\nu \leftrightarrow \beta \right) \right) +\left[
E^{\mu \nu },E^{\alpha \beta }\right]   \tag{32}
\end{equation}

The first term is 
\begin{equation}
\left[ l^{\mu \nu },l^{\alpha \beta }\right] =\left[ \left( x^{\mu }p^{\nu
}-x^{\nu }p^{\mu }\right) ,l^{\alpha \beta }\right]   \tag{33}
\end{equation}

The relation (29) immediately leads us to 
\begin{multline}
\left[ x^{\mu },l^{\alpha \beta }\right] =\imath Q\left( g^{\mu \alpha
}x^{\beta }-g^{\mu \beta }x^{\alpha }\right)   \notag \\
+2\sum_{\sigma _{1}\neq \sigma _{2}}\left( x^{\mu \left( \sigma _{1}\right)
}x^{\alpha \left( \sigma _{2}\right) }p^{\beta }-x^{\mu \left( \sigma
_{1}\right) }x^{\beta \left( \sigma _{2}\right) }p^{\alpha }+x^{\alpha
}x^{\mu \left( \sigma _{1}\right) }p^{\beta \left( \sigma _{2}\right)
}-x^{\beta }x^{\mu \left( \sigma _{1}\right) }p^{\alpha \left( \sigma
_{2}\right) }\right)   \tag{34}
\end{multline}

The relation (33), together with (29) and (34) lead us to 
\begin{equation}
\left[ l^{\mu \nu },l^{\alpha \beta }\right] =\imath Q\left( g^{\mu \beta
}l^{\alpha \nu }-g^{\mu \alpha }l^{\beta \nu }+g^{\nu \alpha }l^{\beta \mu
}-g^{\nu \beta }l^{\alpha \mu }\right) +2\sum_{\sigma _{1}\neq \sigma
_{2}}\sum_{\sigma _{3},\sigma _{4}}\mathcal{E}_{\sigma _{1}\sigma _{2}\sigma
_{3}\sigma _{4}}^{\mu \nu \alpha \beta }  \tag{35}
\end{equation}
with 
\begin{multline}
\mathcal{E}_{\sigma _{1}\sigma _{2}\sigma _{3}\sigma _{4}}^{\mu \nu \alpha
\beta }=\left\{ \left[ x^{\mu \left( \sigma _{3}\right) }x^{\alpha \left(
\sigma _{4}\right) }p^{\nu \left( \sigma _{1}\right) }p^{\beta \left( \sigma
_{2}\right) }+x^{\mu \left( \sigma _{3}\right) }p^{\nu \left( \sigma
_{1}\right) }x^{\alpha \left( \sigma _{2}\right) }p^{\beta \left( \sigma
_{4}\right) }+x^{\mu \left( \sigma _{1}\right) }x^{\alpha \left( \sigma
_{2}\right) }p^{\beta \left( \sigma _{3}\right) }p^{\nu \left( \sigma
_{4}\right) }\right. \right.   \notag \\
\left. \left. +x^{\alpha \left( \sigma _{3}\right) }x^{\mu \left( \sigma
_{1}\right) }p^{\beta \left( \sigma _{2}\right) }p^{\nu \left( \sigma
_{4}\right) }\right] -\left( \mu \leftrightarrow \nu ,\alpha ,\beta \right)
-\left( \mu ,\nu ,\alpha \leftrightarrow \beta \right) +\left( \mu
\leftrightarrow \nu ,\alpha \leftrightarrow \beta \right) \right\}   \tag{36}
\end{multline}

Similarly, with the use of (21) one can perform the other commutators of
(32) and find : 
\begin{equation}
\left[ l^{\mu \nu },E^{\alpha \beta }\right] =-2\imath \sum_{\sigma _{1}\neq
\sigma _{2}}\sum_{\sigma _{3},\sigma _{4}}\sum_{n=1}^{+\infty }\frac{1}{n}%
\Theta _{n,\sigma _{1}\sigma _{2}\sigma _{3}\sigma _{4}}^{\mu \nu \alpha
\beta }  \tag{37}
\end{equation}
with 
\begin{equation}
\Theta _{n,\sigma _{1}\sigma _{2}\sigma _{3}\sigma _{4}}^{\mu \nu \alpha
\beta }=\mathcal{E}_{\sigma _{1}\sigma _{2}\sigma _{3}\sigma _{4}}^{\mu \nu
\alpha \beta }\left( x^{\alpha }\longrightarrow \alpha _{-n}^{\alpha
}\,,\,p^{\beta }\longrightarrow \alpha _{n}^{\beta }\right)   \tag{38}
\end{equation}
and 
\begin{multline}
\left[ E^{\mu \nu },E^{\alpha \beta }\right] =+\imath Q\left( g^{\mu \beta
}E^{\alpha \nu }-g^{\mu \alpha }E^{\beta \nu }+g^{\nu \alpha }E^{\beta \mu
}-g^{\nu \beta }E^{\alpha \mu }\right)   \notag \\
-2\sum_{\sigma _{1}\neq \sigma _{2}}\sum_{\sigma _{3},\sigma
_{4}}\sum_{n,m=1}^{+\infty }\frac{1}{nm}K_{n,m,\sigma _{1}\sigma _{2}\sigma
_{3}\sigma _{4}}^{\mu \nu \alpha \beta }  \tag{39}
\end{multline}
where 
\begin{equation}
K_{n,m,\sigma _{1}\sigma _{2}\sigma _{3}\sigma _{4}}^{\mu \nu \alpha \beta }=%
\mathcal{E}_{\sigma _{1}\sigma _{2}\sigma _{3}\sigma _{4}}^{\mu \nu \alpha
\beta }\left( x^{\mu }\longrightarrow \alpha _{-n}^{\mu }\,,\,p^{\nu
}\longrightarrow \alpha _{n}^{\nu }\,,\,x^{\alpha }\longrightarrow \alpha
_{-m}^{\alpha }\,,\,p^{\beta }\longrightarrow \alpha _{m}^{\beta }\right)  
\tag{40}
\end{equation}

The substitution of (36), (37) and (39) in (32) gives 
\begin{multline}
\left[ M^{\mu \nu },M^{\alpha \beta }\right] =\imath Q\left( g^{\mu \beta
}M^{\alpha \nu }-g^{\mu \alpha }M^{\beta \nu }+g^{\nu \alpha }M^{\beta \mu
}-g^{\nu \beta }M^{\alpha \mu }\right)   \notag \\
+2\sum_{\sigma _{1}\neq \sigma _{2}}\left\{ \sum_{\sigma _{3},\sigma _{4}}
\left[ \mathcal{E}_{\sigma _{1}\sigma _{2}\sigma _{3}\sigma _{4}}^{\mu \nu
\alpha \beta }+\imath \sum_{n=1}^{+\infty }\frac{1}{n}\left( \Theta
_{n,\sigma _{1}\sigma _{2}\sigma _{3}\sigma _{4}}^{\alpha \beta \mu \nu
}-\Theta _{n,\sigma _{1}\sigma _{2}\sigma _{3}\sigma _{4}}^{\mu \nu \alpha
\beta }\right) \right. \right.   \notag \\
\left. \left. -\sum_{n,m=1}^{+\infty }\frac{1}{nm}K_{n,m,\sigma _{1}\sigma
_{2}\sigma _{3}\sigma _{4}}^{\mu \nu \alpha \beta }\right] \right\}  
\tag{41}
\end{multline}

Clearly, such unfamiliar results in (27), (31) and (41) are attributable to
the fact that if the Green indices are different, the bilinear relations
between the Green components become anomalous. We thus conclude that the
Poincar\'{e} algebra is violated. It should be noted here that, on one hand,
if we set $Q=1$ (which correspond to the ordinary case ), the Poincar\'{e}
algebra become satisfyed ( indeed, $Q=1$ means that all the terms which
contain a sommation of the type $\sum_{\sigma _{1}\neq \sigma _{2}}$will be
eliminated ), and on the other hand, to perform these commutators we must
use the Green decomposition.

\subsection{Possible calculation methods}

We begin by remarking that, if in Quantum Mechanics, the correspondance
principle . 
\begin{equation*}
QM:\quad \quad \left( x^{\mu }p^{\nu }-p^{\nu }x^{\mu }\right)
\longrightarrow \left( x_{op}^{\mu }p_{op}^{\nu }-x_{op}^{\nu }p_{op}^{\mu
}\right) \ 
\end{equation*}
doesn't cause any order ambiguity problem, it is clearly not the case for
the Paraquantum Mechanics.This leads us to generalize it by the
correspondance 
\begin{equation*}
PQM:\quad \quad \left( x^{\mu }p^{\nu }-p^{\nu }x^{\mu }\right)
\longrightarrow \frac{1}{2}\left\{ \left[ x_{op}^{\mu },p_{op}^{\nu }\right]
_{+}-\left[ x_{op}^{\nu },p_{op}^{\mu }\right] _{+}\right\}
\end{equation*}

We then rewrite the generators $M^{\mu \nu }$basing on an adequate
symetrization which takes the form $M^{\mu \nu }=l^{\mu \nu }+E^{\mu \nu }$
with : 
\begin{equation}
l^{\mu \nu }=\frac{1}{2}\left[ x^{\mu },p^{\nu }\right] _{+}-\left[ x^{\nu
},p^{\mu }\right] _{+}  \tag{42}
\end{equation}
and 
\begin{equation}
E^{\mu \nu }=-\frac{\imath }{2}\sum_{n=1}^{\infty }\frac{1}{n}\left( \left[
\alpha _{-n}^{\mu },\alpha _{n}^{\nu }\right] _{+}-\left[ \alpha _{-n}^{\nu
},\alpha _{n}^{\mu }\right] _{+}\right)   \tag{43}
\end{equation}

Now, if this writing allows the elimination of the order ambiguities, it
also allows the paraquantum treatment of the problem solely with trilinear
relations (11-18) without having recourse to the Green representation (21).
One can nevertheless find again the same results with the use of the Green
decomposition.

\subsubsection{Direct application of the trilinear relations:}

Let us perform the second commutator of the Poincar\'{e} algebra 
\begin{multline}
\left[ p^{\mu },M^{\nu \rho }\right] =\left[ p^{\mu },\left\{ \frac{1}{2}%
\left[ x^{\nu },p^{\rho }\right] _{+}-\left[ x^{\rho },p^{\nu }\right]
_{+}\right\} \right]   \notag \\
-\imath \sum_{n=1}^{\infty }\frac{1}{n}\left[ p^{\mu },\left\{ \left[ \alpha
_{-n}^{\nu },\alpha _{n}^{\rho }\right] _{+}-\left[ \alpha _{-n}^{\rho
},\alpha _{n}^{\nu }\right] _{+}\right\} \right]   \tag{44}
\end{multline}

With the use of (14), (44) gives 
\begin{equation}
\left[ p^{\mu },M^{\nu \rho }\right] =-\imath g^{\mu \nu }p^{\rho }+\imath
g^{\mu \rho }p^{\nu }  \tag{45}
\end{equation}
a result which satisfy Poincar\'{e} algebra. Similarly, for the third
commutator of the algebra, one can write : 
\begin{equation}
\left[ M^{\mu \nu },M^{\rho \sigma }\right] =\left[ l^{\mu \nu },l^{\rho
\sigma }\right] +\left[ E^{\mu \nu },E^{\rho \sigma }\right] +\left[ l^{\mu
\nu },E^{\rho \sigma }\right] +\left[ E^{\mu \nu },l^{\rho \sigma }\right]  
\tag{46}
\end{equation}

With the direct use of the trilinear relations (11), (14), one can perform
the first term and find (Appendix A ) 
\begin{equation}
\left[ l^{\mu \nu },l^{\rho \sigma }\right] =\imath g^{\nu \rho }l^{\sigma
\mu }-\imath g^{\mu \sigma }l^{\nu \rho }-\imath g^{\nu \sigma }l^{\rho \mu
}-\imath g^{\mu \rho }l^{\sigma \nu }  \tag{47}
\end{equation}

In the same way, one can perform the second term and obtain (Appendix B) 
\begin{equation}
\left[ E^{\mu \nu },E^{\rho \sigma }\right] =\imath \left( g^{\nu \rho
}E^{\sigma \mu }+g^{\mu \rho }E^{\nu \sigma }+g^{\nu \sigma }E^{\mu \rho
}+g^{\mu \sigma }E^{\rho \nu }\right)   \tag{48}
\end{equation}

\bigskip Now the relation $\left[ x^{\mu }p^{\nu },\left[ \alpha _{-n}^{\rho
},\alpha _{n}^{\sigma }\right] _{+}\right] =0$ , leads to the result $\left[
l^{\mu \nu },E^{\rho \sigma }\right] =\left[ E^{\mu \nu },l^{\rho \sigma }%
\right] =0$

When combined with (47) and (48), (46) gives

\begin{equation}
\left[ M^{\mu \nu },M^{\rho \sigma }\right] =\imath g^{\nu \rho }M^{\sigma
\mu }-\imath g^{\mu \sigma }M^{\nu \rho }-\imath g^{\nu \sigma }M^{\rho \mu
}+\imath g^{\mu \rho }M^{\nu \sigma }  \tag{49}
\end{equation}

\subsubsection{Green decomposition}

In terms of Green components, the generators $M^{\mu \nu }$ are given by : 
\begin{multline}
M^{\mu \nu }=\frac{1}{2}\sum_{\sigma _{1}=1}^{Q}\sum_{\sigma
_{2}=1}^{Q}\left\{ \left[ x^{\mu \left( \sigma _{1}\right) },p^{\nu \left(
\sigma _{2}\right) }\right] _{+}-\left[ x^{\nu \left( \sigma _{1}\right)
},p^{\mu \left( \sigma _{2}\right) }\right] _{+}\right.   \notag \\
\left. -\imath \sum_{n=1}^{\infty }\frac{1}{n}\left( \left[ \alpha
_{-n}^{\mu \left( \sigma _{1}\right) },\alpha _{n}^{\nu \left( \sigma
_{2}\right) }\right] _{+}-\left[ \alpha _{-n}^{\nu \left( \sigma _{1}\right)
},\alpha _{n}^{\mu \left( \sigma _{2}\right) }\right] _{+}\right) \right\}  
\tag{50}
\end{multline}
From the relations (21), (50) can be rewritten as : 
\begin{equation}
M^{\mu \nu }=\sum_{\sigma =1}^{Q}\left\{ x^{\mu \left( \sigma \right)
}p^{\nu \left( \sigma \right) }-x^{\nu \left( \sigma \right) }p^{\mu \left(
\sigma \right) }-\imath \sum_{n=1}^{\infty }\frac{1}{n}\left( \alpha
_{-n}^{\mu \left( \sigma \right) }\alpha _{n}^{\nu \left( \sigma \right)
}-\alpha _{-n}^{\nu \left( \sigma \right) }\alpha _{n}^{\mu \left( \sigma
\right) }\right) \right\}   \tag{51}
\end{equation}
Then, the symetrization of $M^{\mu \nu }$ allows us to obtain the result : 
\begin{equation*}
M^{\mu \nu }=\sum_{\sigma _{1},\sigma _{2}=1}^{Q}M^{\mu \nu \left( \sigma
_{1},\sigma _{2}\right) }=\sum_{\sigma =1}^{Q}M^{\mu \nu \left( \sigma
\right) }
\end{equation*}

\bigskip where $M^{\mu \nu (\sigma )}$ has the ordinary form.

One can then perform the two commutators $\left[ p^{\mu },M^{\nu \rho }%
\right] $ and $\left[ M^{\mu \nu },M^{\rho \sigma }\right] $ as follows : 
\begin{eqnarray}
\left[ p^{\mu },M^{\nu \rho }\right]  &=&\sum_{\sigma _{1},\sigma _{2}=1}^{Q}%
\left[ p^{\mu \left( \sigma _{1}\right) },M^{\nu \rho \left( \sigma
_{2}\right) }\right]   \notag \\
&=&\sum_{\sigma =1}^{Q}\left[ p^{\mu \left( \sigma \right) },M^{\nu \rho
\left( \sigma \right) }\right] +\sum_{\sigma _{1}\neq \sigma _{2}}\left[
p^{\mu \left( \sigma _{1}\right) },M^{\nu \rho \left( \sigma _{2}\right) }%
\right]   \TCItag{52}
\end{eqnarray}
By the use of 
\begin{equation}
\left[ A^{\left( \sigma _{1}\right) },B^{\left( \sigma _{2}\right)
}.C^{\left( \sigma _{2}\right) }\right] =0\quad \quad \quad \forall
A,B,C\,\,\,\text{for \thinspace }\sigma _{1}\neq \sigma _{2}  \tag{53}
\end{equation}
It is clear that 
\begin{equation}
\left[ p^{\mu \left( \sigma _{1}\right) },M^{\nu \rho \left( \sigma
_{2}\right) }\right] =0\quad \quad \quad \sigma _{1}\neq \sigma _{2} 
\tag{54}
\end{equation}
and 
\begin{eqnarray}
\left[ p^{\mu },M^{\nu \rho }\right]  &=&\sum_{\sigma =1}^{Q}\left[ p^{\mu
\left( \sigma \right) },M^{\nu \rho \left( \sigma \right) }\right]   \notag
\\
&=&\sum_{\sigma =1}^{Q}\left( \imath g^{\mu \rho }p^{\nu \left( \sigma
\right) }-\imath g^{\mu \nu }p^{\rho \left( \sigma \right) }\right)  
\TCItag{55}
\end{eqnarray}
Then 
\begin{equation}
\left[ p^{\mu },M^{\nu \rho }\right] =\imath g^{\mu \rho }p^{\nu }-\imath
g^{\mu \nu }p^{\rho }  \tag{56}
\end{equation}
In the same way 
\begin{equation}
\left[ M^{\mu \nu },M^{\rho \sigma }\right] =\sum_{\alpha ,\beta =1}^{Q}%
\left[ M^{\mu \nu \left( \alpha \right) },M^{\rho \sigma \left( \beta
\right) }\right]   \tag{57}
\end{equation}
But 
\begin{equation}
\left[ A^{\left( \alpha \right) }B^{\left( \alpha \right) },C^{\left( \beta
\right) }D^{\left( \beta \right) }\right] =0\quad \quad \quad \alpha \neq
\beta   \tag{58}
\end{equation}
Then 
\begin{eqnarray}
\left[ M^{\mu \nu },M^{\rho \sigma }\right]  &=&\sum_{\alpha =1}^{Q}\left[
M^{\mu \nu \left( \alpha \right) },M^{\rho \sigma \left( \alpha \right) }%
\right]   \notag \\
&=&\sum_{\alpha =1}^{Q}\left( \imath g^{\nu \rho }M^{\sigma \mu \left(
\alpha \right) }-\imath g^{\mu \sigma }M^{\nu \rho \left( \alpha \right)
}-\imath g^{\nu \sigma }M^{\rho \mu \left( \alpha \right) }+\imath g^{\mu
\sigma }M^{\rho \nu \left( \alpha \right) }\right)   \TCItag{59}
\end{eqnarray}
Lastly 
\begin{equation}
\left[ M^{\mu \nu },M^{\rho \sigma }\right] =\imath g^{\nu \rho }M^{\sigma
\mu }-\imath g^{\mu \sigma }M^{\nu \rho }-\imath g^{\nu \sigma }M^{\rho \mu
}+\imath g^{\mu \sigma }M^{\rho \nu }  \tag{60}
\end{equation}

\section{Space-time critical dimensions}

Let us introduce, in the transverse gauge, the generators $M^{i-}$ in the
form : 
\begin{equation}
M^{i-}=l^{i-}+E^{i-}  \tag{61}
\end{equation}
where 
\begin{equation}
l^{i-}=\frac{1}{2}\left[ x^{i},\frac{1}{p^{+}}\right] _{+}\alpha _{0}^{-}-%
\frac{1}{2}\left[ x^{-},p^{i}\right] _{+}  \tag{62}
\end{equation}
and 
\begin{equation}
E^{i-}=-\frac{\imath }{2}\sum_{n=1}^{\infty }\frac{1}{n}\left( \left[ \alpha
_{-n}^{i},\frac{1}{p^{+}}\right] _{+}\alpha _{n}^{-}-\alpha _{-n}^{-}\left[
\alpha _{n}^{i},\frac{1}{p^{+}}\right] _{+}\right)   \tag{63}
\end{equation}
One can write 
\begin{equation}
\left[ M^{i-},M^{j-}\right] =\left[ l^{i-},l^{j-}\right] +\left[
l^{i-},E^{j-}\right] +\left[ E^{i-},l^{j-}\right] +\left[ E^{i-},E^{j-}%
\right]   \tag{64}
\end{equation}
It is easy to verify that: 
\begin{equation}
\left[ \alpha ^{i},\alpha _{0}^{-}\right] =-\imath p^{i}  \tag{65}
\end{equation}
and 
\begin{equation}
\left[ x^{i},\alpha _{n}^{-}\right] =-\imath \alpha _{n}^{i}  \tag{66}
\end{equation}
By the use of the trilinear relations (22) and the Heisenberg equations, one
can verify that : 
\begin{multline}
\left[ l^{i-},l^{j-}\right] =\frac{\imath }{4}\left( \left[ x^{i},\frac{1}{%
p^{+}}\right] _{+}\left[ p^{j},\frac{1}{p^{+}}\right] _{+}-\left[ x^{j},%
\frac{1}{p^{+}}\right] _{+}\left[ p^{i},\frac{1}{p^{+}}\right] _{+}\right)
\alpha _{0}^{-}  \notag \\
-\frac{\imath }{2}\left( \frac{1}{\left( p^{+}\right) ^{2}}\left[ x^{i},p^{j}%
\right] _{+}-\frac{1}{\left( p^{+}\right) ^{2}}\left[ x^{j},p^{i}\right]
_{+}\right) \alpha _{0}^{-}  \tag{67}
\end{multline}
and 
\begin{multline}
\left[ l^{i-},E^{j-}\right] =\frac{\imath }{4}\sum_{n}\frac{1}{n}\left\{
\left( \left[ \alpha _{-n}^{i},\frac{1}{p^{+}}\right] _{+}\left[ \alpha
_{n}^{j},\frac{1}{p^{+}}\right] _{+}-\left[ \alpha _{-n}^{j},\frac{1}{p^{+}}%
\right] _{+}\left[ \alpha _{n}^{i},\frac{1}{p^{+}}\right] _{+}\right) \alpha
_{0}^{-}\right.   \notag \\
\left. +\frac{2}{\left( p^{+}\right) ^{2}}\left[ p^{i},\alpha _{-n}^{j}%
\right] _{+}\alpha _{n}^{-}-\frac{2}{\left( p^{+}\right) ^{2}}\alpha
_{-n}^{-}\left[ p^{i},\alpha _{n}^{j}\right] _{+}\right\}   \tag{68}
\end{multline}
If we take the mean values on the physical states $\left( \alpha
_{-m}^{k}\left| 0\right\rangle \right) $ and by the use of the trilinear
relations (22), one can prove that : 
\begin{equation}
\left\langle 0\right| \alpha _{m}^{l}\left[ l^{i-},l^{j-}\right] \alpha
_{-m}^{k}\left| 0\right\rangle =0  \tag{69}
\end{equation}
\begin{equation}
\left\langle 0\right| \alpha _{m}^{l}\left[ l^{i-},E^{j-}\right] \alpha
_{-m}^{k}\left| 0\right\rangle =\frac{m^{2}}{\left( p^{+}\right) ^{2}}\left(
\delta ^{li}\delta ^{kj}-\delta ^{lj}\delta ^{ki}\right) +\frac{m}{\left(
p^{+}\right) ^{2}}\left( \delta ^{lj}p^{i}p^{k}-\delta
^{jk}p^{l}p^{i}\right)   \tag{70}
\end{equation}
Finally 
\begin{multline}
\left\langle 0\right| \alpha _{m}^{l}\left( \left[ l^{i-},E^{j-}\right] +%
\left[ E^{i-},l^{j-}\right] \right) \alpha _{-m}^{k}\left| 0\right\rangle
=2m^{2}\left( \frac{1}{p^{+}}\right) ^{2}\left( \delta ^{li}\delta
^{kj}-\delta ^{lj}\delta ^{ki}\right)   \notag \\
+\frac{m}{\left( p^{+}\right) ^{2}}\left( \delta ^{lj}p^{i}p^{k}-\delta
^{li}p^{j}p^{k}-\delta ^{jk}p^{l}p^{i}+\delta ^{ik}p^{l}p^{j}\right)  
\tag{71}
\end{multline}
On the other hand, by the use of the trilinear relations (22), one can write
: 
\begin{equation}
\left\langle 0\right| \alpha _{m}^{l}E^{i-}E^{j-}\alpha _{-m}^{k}\left|
0\right\rangle =\sum_{i=1}^{4}C_{i}  \tag{72}
\end{equation}
Where 
\begin{eqnarray}
C_{1} &=&-\frac{1}{4}\sum_{n,n^{\prime }}\left\langle 0\right| \alpha
_{m}^{l}\left[ \frac{\alpha _{-n}^{i}}{n},\frac{1}{p^{+}}\right] _{+}\alpha
_{n}^{-}\left[ \frac{\alpha _{-n^{\prime }}^{j}}{n^{\prime }},\frac{1}{p^{+}}%
\right] _{+}\alpha _{n^{\prime }}^{-}\alpha _{-m}^{k}\left| 0\right\rangle  
\notag \\
&=&-\frac{m}{p^{+}}\delta ^{li}\left( \frac{1}{2}m\left( m-1\right) \frac{%
\delta ^{jk}}{p^{+}}+\frac{1}{2}\left[ p^{j},\frac{1}{p^{+}}\right]
_{+}p^{k}\right)   \TCItag{73}
\end{eqnarray}
\begin{eqnarray}
C_{2} &=&\frac{1}{4}\sum_{n,n^{\prime }}\left\langle 0\right| \alpha
_{m}^{l}\alpha _{-n}^{-}\left[ \frac{\alpha _{n}^{i}}{n},\frac{1}{p^{+}}%
\right] _{+}\left[ \frac{\alpha _{-n^{\prime }}^{j}}{n^{\prime }},\frac{1}{%
p^{+}}\right] _{+}\alpha _{n^{\prime }}^{-}\alpha _{-m}^{k}\left|
0\right\rangle   \notag \\
&=&-\frac{m^{2}}{\left( p^{+}\right) ^{2}}\left( m-1\right) \delta
^{lj}\delta ^{ik}  \TCItag{74}
\end{eqnarray}
\begin{eqnarray}
C_{3} &=&\frac{1}{4}\sum_{n,n^{\prime }}\left\langle 0\right| \alpha
_{m}^{l}\alpha _{-n}^{-}\left[ \frac{\alpha _{-n}^{i}}{n},\frac{1}{p^{+}}%
\right] _{+}\alpha _{-n^{\prime }}^{-}\left[ \frac{\alpha _{-n^{\prime }}^{j}%
}{n^{\prime }},\frac{1}{p^{+}}\right] _{+}\alpha _{-m}^{k}\left|
0\right\rangle   \notag \\
&=&\frac{m}{p^{+}}\delta ^{jk}\left( m\frac{\delta ^{li}}{2p^{+}}\left(
m-1\right) +p^{l}p^{i}\right)   \TCItag{75}
\end{eqnarray}
\begin{eqnarray}
C_{4} &=&\frac{1}{4}\sum_{n,n^{\prime }}\left\langle 0\right| \alpha _{m}^{l}%
\left[ \frac{\alpha _{-n}^{i}}{n},\frac{1}{p^{+}}\right] _{+}\alpha
_{-n}^{-}\alpha _{-n^{\prime }}^{-}\left[ \frac{\alpha _{-n^{\prime }}^{j}}{%
n^{\prime }},\frac{1}{p^{+}}\right] _{+}\alpha _{-m}^{k}\left|
0\right\rangle   \notag \\
&=&\frac{1}{\left( p^{+}\right) ^{2}}\delta ^{li}\delta ^{jk}\left( Q\frac{%
D-2}{12}m\left( m^{2}-1\right) +2m\alpha \left( 0\right) \right)  
\TCItag{76}
\end{eqnarray}
Lastly 
\begin{multline}
\left\langle 0\right| \alpha _{m}^{l}E^{i-}E^{j-}\alpha _{-m}^{k}\left|
0\right\rangle =-\left( \frac{1}{p^{+}}\right) ^{2}m^{2}\left( m-1\right)
\left( \delta ^{li}\delta ^{jk}-\delta ^{lj}\delta ^{ik}\right)   \notag \\
+\left( \frac{1}{p^{+}}\right) ^{2}\delta ^{li}\delta ^{jk}\left( Q\frac{D-2%
}{12}m\left( m^{2}-1\right) +2m\alpha \left( 0\right) \right) +\frac{m}{%
\left( p^{+}\right) ^{2}}\left( \delta ^{jk}p^{l}p^{i}-\delta
^{li}p^{j}p^{k}\right)   \tag{77}
\end{multline}
Then 
\begin{multline}
\left\langle 0\right| \alpha _{m}^{l}\left[ E^{i-},E^{j-}\right] \alpha
_{-m}^{k}\left| 0\right\rangle =  \notag \\
-\left( \frac{1}{p^{+}}\right) ^{2}\left( \delta ^{li}\delta ^{jk}-\delta
^{lj}\delta ^{ik}\right) \left( -2m^{2}\left( m-1\right) +Q\frac{D-2}{12}%
m\left( m^{2}-1\right) +2m\alpha \left( 0\right) \right)   \notag \\
-\frac{m}{\left( p^{+}\right) ^{2}}\left( \delta ^{li}p^{j}p^{k}-\delta
^{lj}p^{i}p^{k}-\delta ^{kj}p^{l}p^{i}+\delta ^{ik}p^{l}p^{i}\right)  
\tag{78}
\end{multline}
Collecting these results, one obtain : 
\begin{multline}
\left\langle 0\right| \alpha _{m}^{l}\left[ M^{i-},M^{j-}\right] \alpha
_{-m}^{k}\left| 0\right\rangle =  \notag \\
-\left( \frac{1}{p^{+}}\right) ^{2}\left( \delta ^{li}\delta ^{jk}-\delta
^{lj}\delta ^{ik}\right) \left( -2m^{3}+Q\frac{D-2}{12}m\left(
m^{2}-1\right) +2m\alpha \left( 0\right) \right)   \tag{79}
\end{multline}
In terms of operators, this equation is equivalent to : 
\begin{multline}
\left[ M^{i-},M^{j-}\right] =  \notag \\
-\frac{1}{2\left( p^{+}\right) ^{2}}\sum_{n=1}^{\infty }\left( \left[ \alpha
_{-n}^{i},\alpha _{n}^{j}\right] _{+}-\left[ \alpha _{-n}^{j},\alpha _{n}^{i}%
\right] _{+}\right) \left( -2n+Q\frac{D-2}{12}\left( n-\frac{1}{n}\right)
+2\alpha \left( 0\right) .\frac{1}{n}\right)   \tag{80}
\end{multline}
In conclusion, to have $\left( \left[ M^{i-},M^{j-}\right] =0\right) ,$
equation (80) gives 
\begin{equation}
\left\{ 
\begin{array}{c}
\displaystyle D=2+\frac{24}{Q} \\ 
\alpha \left( 0\right) =1
\end{array}
\right.   \tag{81}
\end{equation}

\section{Conclusion}

In view of the importance of the Poincar\'{e} algebra , the aim of this work
is to study the consequences which result from the paraquantization of this
algebra in string theory. Unlike Ardalan and Mansouri work [6], in this
study, the paraquantization of the theory requires the same treatment for
the center of mass variables and the excitation modes of the string. The
first consequence is the appearance of noncommuting momentum coordinates
expressed by the relation (27) which is equivalent to $\left[ p^{\mu },\left[
p^{\nu },p^{\sigma }\right] _{+}\right] =0.$ For the two other commutators
of the algebra, the generalization of the correspondance principle permits
to redefine the generators $M^{\mu \nu }$ in both Lorentz and transverse
gauges. As a concequence, one can find again the results :

\begin{equation*}
\left[ p^{\mu },M^{\nu \rho }\right] =-\imath g^{\mu \nu }p^{\rho }+\imath
g^{\mu \rho }p^{\nu }
\end{equation*}
\begin{equation*}
\left[ M^{\mu \nu },M^{\rho \sigma }\right] =\imath g^{\nu \rho }M^{\sigma
\mu }-\imath g^{\mu \sigma }M^{\nu \rho }-\imath g^{\nu \sigma }M^{\rho \mu
}+\imath g^{\mu \rho }M^{\nu \sigma }
\end{equation*}
\ with the use of the trilinear relations or the Green decomposition. In the
transverse gauge, to re-establish the commutator $\left( \left[ M^{i-},M^{j-}%
\right] =0\right) $, the space-time critical dimension $D$ is calculated
with the only use of the trilinear relations. Like in Ardalan and Mansouri
work [6], $D$ is again given as a function of the paraquantization order $Q$
through the relation $\displaystyle D=2+\frac{24}{Q}$. This result opens
other existence possibilities of bosonic strings at space-time dimensions $%
D=26,14,10,8,6,5,4$ respectively in orders $Q=1,2,3,4,6,8,12$. Notice here
that in terms of Green components one gets $\left[ x^{\mu },x^{\nu }\right]
=2\sum_{\alpha \neq \beta }x^{\mu \left( \alpha \right) }x^{\nu \left( \beta
\right) }\neq 0\,!$, that is noncommuting coordinates. One may then wonder
if this is a kind of noncommuting space-time, since we have in addition
noncommuting momentum coordinates.

\newpage\textbf{\Large Appendices}

\appendix

\section{\protect\bigskip}

One can write 
\begin{eqnarray}
\left[ l^{\mu \nu },l^{\rho \sigma }\right]  &=&\frac{1}{4}\left\{ \left[ %
\left[ x^{\mu },p^{\nu }\right] _{+},\left[ x^{\rho },p^{\sigma }\right] _{+}%
\right] -\left[ \left( \nu ,\mu \right) ,\left( \rho ,\sigma \right) \right]
-\left[ \left( \mu ,\nu \right) ,\left( \sigma ,\rho \right) \right] +\left[
\left( \nu ,\mu \right) ,\left( \sigma ,\rho \right) \right] \right\}  
\notag \\
&=&\frac{1}{4}\left( A-B-C+D\right)   \TCItag{82}
\end{eqnarray}
\bigskip Where 
\begin{equation}
A=x^{\mu }\left[ p^{\nu },\left[ x^{\rho },p^{\sigma }\right] _{+}\right] +%
\left[ x^{\mu },\left[ x^{\rho },p^{\sigma }\right] _{+}\right] p^{\nu
}+p^{\nu }\left[ x^{\mu },\left[ x^{\rho },p^{\sigma }\right] _{+}\right] +%
\left[ p^{\nu },\left[ x^{\rho },p^{\sigma }\right] _{+}\right] x^{\mu } 
\tag{83}
\end{equation}
\bigskip by the use of (11), (14),  (83) gives 
\begin{equation}
A=-2\imath g^{\nu \rho }\left[ x^{\mu },p^{\sigma }\right] _{+}+2\imath
g^{\mu \sigma }\left[ x^{\rho },p^{\nu }\right] _{+}  \tag{84}
\end{equation}

In the same way, one can compute

\begin{eqnarray}
B &=&A\left( \mu \leftrightarrow \nu \right) =-2\imath g^{\mu \rho }\left[
x^{\nu },p^{\sigma }\right] _{+}+2\imath g^{\nu \sigma }\left[ x^{\rho
},p^{\mu }\right] _{+}  \TCItag{85} \\
C &=&A\left( \rho \leftrightarrow \sigma \right) =-2\imath g^{\nu \sigma }
\left[ x^{\mu },p^{\rho }\right] _{+}+2\imath g^{\mu \rho }\left[ x^{\sigma
},p^{\nu }\right] _{+}  \TCItag{86} \\
D &=&A\left( \mu \leftrightarrow \nu ,\rho \leftrightarrow \sigma \right)
=-2\imath g^{\mu \sigma }\left[ x^{\nu },p^{\rho }\right] _{+}+2\imath
g^{\nu \rho }\left[ x^{\sigma },p^{\mu }\right] _{+}  \TCItag{87}
\end{eqnarray}

By substitution in (82), one obtain :

\begin{eqnarray}
\left[ l^{\mu \nu },l^{\rho \sigma }\right]  &=&\frac{\imath }{2}g^{\nu \rho
}\left( \left[ x^{\sigma },p^{\mu }\right] _{+}-\left[ x^{\mu },p^{\sigma }%
\right] _{+}\right) +\frac{\imath }{2}g^{\mu \sigma }\left( \left[ x^{\rho
},p^{\nu }\right] _{+}-\left[ x^{\nu },p^{\rho }\right] _{+}\right)   \notag
\\
&&-\frac{\imath }{2}g^{\mu \rho }\left( \left[ x^{\sigma },p^{\nu }\right]
_{+}-\left[ x^{\nu },p^{\sigma }\right] _{+}\right) -\frac{\imath }{2}g^{\nu
\sigma }\left( \left[ x^{\rho },p^{\mu }\right] _{+}-\left[ x^{\mu },p^{\rho
}\right] _{+}\right)   \TCItag{88}
\end{eqnarray}
Then 
\begin{equation}
\left[ l^{\mu \nu },l^{\rho \sigma }\right] =\imath g^{\nu \rho }l^{\sigma
\mu }-\imath g^{\mu \sigma }l^{\nu \rho }-\imath g^{\nu \sigma }l^{\rho \mu
}-\imath g^{\mu \rho }l^{\sigma \nu }  \tag{89}
\end{equation}

\section{\protect\bigskip}

One can write 
\begin{eqnarray}
\left[ E^{\mu \nu },E^{\rho \sigma }\right]  &=&-\frac{1}{4}%
\sum_{n=1}^{\infty }\sum_{m=1}^{\infty }\frac{1}{nm}\left\{ \left[ \left[
\alpha _{-n}^{\mu },\alpha _{n}^{\nu }\right] _{+},\left[ \alpha _{-m}^{\rho
},\alpha _{m}^{\sigma }\right] _{+}\right] -\left[ \left( \nu ,\mu \right)
,\left( \rho ,\sigma \right) \right] \right.   \notag \\
&&\quad \quad \quad \quad \quad \quad \quad \quad \quad \quad \quad \left. -
\left[ \left( \mu ,\nu \right) ,\left( \sigma ,\rho \right) \right] +\left[
\left( \nu ,\mu \right) ,\left( \sigma ,\rho \right) \right] \right\}  
\notag \\
&=&-\frac{1}{4}\left( A^{\prime }-B^{\prime }-C^{\prime }+D^{\prime }\right) 
\TCItag{90}
\end{eqnarray}

\bigskip where 
\begin{eqnarray}
A^{\prime } &=&\sum_{n,m=1}^{\infty }\frac{1}{nm}\left\{ \alpha _{-n}^{\mu }%
\left[ \alpha _{n}^{\nu },\left[ \alpha _{-m}^{\rho },\alpha _{m}^{\sigma }%
\right] _{+}\right] +\left[ \alpha _{-n}^{\mu },\left[ \alpha _{-m}^{\rho
},\alpha _{m}^{\sigma }\right] _{+}\right] \alpha _{n}^{\nu }\right.   \notag
\\
&&\quad \quad \quad \quad \quad \quad \quad \quad \quad \left. \alpha
_{n}^{\nu }\left[ \alpha _{-n}^{\mu },\left[ \alpha _{-m}^{\rho },\alpha
_{m}^{\sigma }\right] _{+}\right] +\left[ \alpha _{n}^{\nu },\left[ \alpha
_{-m}^{\rho },\alpha _{m}^{\sigma }\right] _{+}\right] \alpha _{-n}^{\mu
}\right\}   \TCItag{91}
\end{eqnarray}

By the use of (16), one can write

\begin{eqnarray}
A^{\prime } &=&2\sum_{n,m=1}^{\infty }\frac{1}{nm}\left\{ \alpha _{-n}^{\mu
}\left( g^{\nu \rho }n\delta _{n-m,0}\alpha _{m}^{\sigma }+g^{\nu \sigma
}n\delta _{n+m,0}\alpha _{-m}^{\rho }\right) \right.   \notag \\
&&\quad \quad \quad \quad \quad +\left( g^{\mu \rho }\left( -n\right) \delta
_{-n-m,0}\alpha _{m}^{\sigma }+g^{\mu \sigma }\left( -n\right) \delta
_{-n+m,0}\alpha _{-m}^{\rho }\right) \alpha _{n}^{\nu }  \notag \\
&&\quad \quad \quad \quad \quad +\alpha _{n}^{\nu }\left( g^{\mu \rho
}\left( -n\right) \delta _{-n-m,0}\alpha _{m}^{\sigma }+g^{\mu \sigma
}\left( -n\right) \delta _{-n+m,0}\alpha _{-m}^{\rho }\right)   \notag \\
&&\quad \quad \quad \quad \quad \left. +\left( g^{\nu \rho }n\delta
_{n-m,0}\alpha _{m}^{\sigma }+g^{\nu \sigma }n\delta _{n+m,0}\alpha
_{-m}^{\rho }\right) \alpha _{-n}^{\mu }\right\}   \notag \\
&=&2\sum_{n=1}^{\infty }\frac{1}{n}\left\{ \alpha _{-n}^{\mu }\left( g^{\nu
\rho }\alpha _{n}^{\sigma }\right) -\left( g^{\mu \sigma }\alpha _{-n}^{\rho
}\right) \alpha _{n}^{\nu }-\alpha _{n}^{\nu }\left( g^{\mu \sigma }\alpha
_{-n}^{\rho }\right) +\left( g^{\nu \rho }\alpha _{n}^{\sigma }\right)
\alpha _{-n}^{\mu }\right\}   \TCItag{92}
\end{eqnarray}
Then 
\begin{equation}
A^{\prime }=2\sum_{n=1}^{\infty }\frac{1}{n}\left\{ g^{\nu \rho }\left[
\alpha _{-n}^{\mu },\alpha _{n}^{\sigma }\right] _{+}-g^{\mu \sigma }\left[
\alpha _{-n}^{\rho },\alpha _{n}^{\nu }\right] _{+}\right\}   \tag{93}
\end{equation}
We can then deduce 
\begin{eqnarray}
-B^{\prime } &=&2\sum_{n=1}^{\infty }\frac{1}{n}\left\{ -g^{\mu \rho }\left[
\alpha _{-n}^{\nu },\alpha _{n}^{\sigma }\right] _{+}+g^{\nu \sigma }\left[
\alpha _{-n}^{\rho },\alpha _{n}^{\mu }\right] _{+}\right\}   \TCItag{94} \\
-C^{\prime } &=&2\sum_{n=1}^{\infty }\frac{1}{n}\left\{ -g^{\nu \sigma }%
\left[ \alpha _{-n}^{\mu },\alpha _{n}^{\rho }\right] _{+}+g^{\mu \rho }%
\left[ \alpha _{-n}^{\sigma },\alpha _{n}^{\nu }\right] _{+}\right\}  
\TCItag{95} \\
D^{\prime } &=&2\sum_{n=1}^{\infty }\frac{1}{n}\left\{ g^{\mu \sigma }\left[
\alpha _{-n}^{\nu },\alpha _{n}^{\rho }\right] _{+}-g^{\nu \rho }\left[
\alpha _{-n}^{\sigma },\alpha _{n}^{\mu }\right] _{+}+\right\}   \TCItag{96}
\end{eqnarray}
And finally 
\begin{eqnarray}
\left[ E^{\mu \nu },E^{\rho \sigma }\right]  &=&\imath \left\{ g^{\nu \rho
}\left( \frac{-\imath }{2}\right) \sum_{n=1}^{\infty }\frac{1}{n}\left( %
\left[ \alpha _{-n}^{\sigma },\alpha _{n}^{\mu }\right] _{+}-\left[ \alpha
_{-n}^{\mu },\alpha _{n}^{\sigma }\right] _{+}\right) \right.   \notag \\
&&\quad \quad \quad \quad \quad +g^{\nu \sigma }\left( \frac{-\imath }{2}%
\right) \sum_{n=1}^{\infty }\frac{1}{n}\left( \left[ \alpha _{-n}^{\mu
},\alpha _{n}^{\rho }\right] _{+}-\left[ \alpha _{-n}^{\rho },\alpha
_{n}^{\mu }\right] _{+}\right)   \notag \\
&&\quad \quad \quad \quad \quad +g^{\mu \rho }\left( \frac{-\imath }{2}%
\right) \sum_{n=1}^{\infty }\frac{1}{n}\left( \left[ \alpha _{-n}^{\nu
},\alpha _{n}^{\sigma }\right] _{+}-\left[ \alpha _{-n}^{\sigma },\alpha
_{n}^{\nu }\right] _{+}\right)   \notag \\
&&\quad \quad \quad \quad \,\quad \left. +g^{\mu \sigma }\left( \frac{%
-\imath }{2}\right) \sum_{n=1}^{\infty }\frac{1}{n}\left( \left[ \alpha
_{-n}^{\rho },\alpha _{n}^{\nu }\right] _{+}-\left[ \alpha _{-n}^{\nu
},\alpha _{n}^{\rho }\right] _{+}\right) \right\}   \TCItag{97}
\end{eqnarray}
Which gives 
\begin{equation}
\left[ E^{\mu \nu },E^{\rho \sigma }\right] =\imath \left( g^{\nu \rho
}E^{\sigma \mu }+g^{\mu \rho }E^{\nu \sigma }+g^{\nu \sigma }E^{\mu \rho
}+g^{\mu \sigma }E^{\rho \nu }\right)   \tag{98}
\end{equation}


\begin{thebibliography}{9}
\bibitem{1}  E. P. Wigner, Phys. Rev. 77, (1950) 711.

\bibitem{2}  H. S Green, Phys. Rev. 90, (1953) 270.

\bibitem{3}  Y.Ohnuki, S. Kamefuchi, \textit{Quantum Field Theory and
parastatistics}, Springer-Verlag, 1982.

\bibitem{4}  Y. Nambu, Lectures at the Copenhagen Summer Symposium, (1970),
unpublished.

\bibitem{5}  T.Goto, Prog. Theor. Phys. 46, 1560 (1971).

\bibitem{6}  F. Ardalan and F. Mansouri, Phys.Rev. D9, 3341(1974).

\bibitem{7}  N. Mebarki, M. Haouchine, N. Belaloui, Acta Phys.Pol. B28,
2017(1997).

\bibitem{8}  N. Mebarki, M. Haouchine, N. Belaloui, Acta Phys.Pol. B28,
2007(1997).
\end{thebibliography}
\end{document}